\documentclass[11pt]{article}
\newtheorem{defn} {DEFINITION} 

\usepackage{tabularx}
\usepackage{times}
\usepackage{amsfonts}
\usepackage{graphicx}
\usepackage{amsmath}
\usepackage{ascmac}
\usepackage{amssymb}
\usepackage[ruled,vlined,linesnumbered]{algorithm2e}
\usepackage{booktabs}
\usepackage{bm}
\usepackage{color}
\usepackage{float}
\usepackage{tabularx}
\usepackage{comment}
\usepackage{here}
\usepackage{multirow}
\usepackage{mwe}
\usepackage[export]{adjustbox}
\usepackage[labelfont=bf]{caption}

\usepackage{url}
\title{A Novel Global Spatial Attention Mechanism in Convolutional Neural Network for Medical Image Classification}
\date{\vspace{-3ex}}

\author{Linchuan Xu$^{*1}$, Jun Huang$^2$, Atsushi Nitanda$^2$, Ryo Asaoka$^3$, Kenji Yamanishi$^2$}
\date{%
    $^1$PQ813, Department of Computing\\
    The Hong Kong Polytechnic University\\
    Hung Hom, Kowloon, Hong Kong\\
    linch.xu@polyu.edu.hk\\%
    $^2$Graduate School of Information Science and Technology\\
    The University of Tokyo, Tokyo, Japan\\
    $\{$jun\_huang, nitanda, yamanishi  $\}$@mist.i.u-tokyo.ac.jp\\%
    $^3$Department of Ophthalmology\\
    The University of Tokyo, Tokyo, Japan\\
    rasaoka-tky@umin.ac.jp\\
}

\begin{document}
\setlength{\abovedisplayskip}{2pt}
\setlength{\belowdisplayskip}{2pt}

\maketitle
\begin{abstract}
Spatial attention has been introduced to convolutional neural networks (CNNs) for improving both their performance and interpretability in visual tasks including image classification. The essence of the spatial attention is to learn a weight map which represents the relative importance of activations within the same layer or channel. All existing attention mechanisms are local attentions in the sense that weight maps are image-specific. However, in the medical field, there are cases that all the images should share the same weight map because the set of images record the same kind of symptom related to the same object and thereby share the same structural content. 
In this paper, we thus propose a novel global spatial attention mechanism in CNNs mainly for medical image classification.
The global weight map is instantiated by a decision boundary between important pixels and unimportant pixels. And we propose to realize the decision boundary by a binary classifier in which the intensities of all images at a pixel are the features of the pixel. The binary classification is integrated into an image classification CNN and is to be optimized together with the CNN. Experiments on two medical image datasets and one facial expression dataset showed that with the proposed attention, not only the performance of four powerful CNNs which are GoogleNet, VGG, ResNet, and DenseNet can be improved, but also meaningful attended regions can be obtained, which is beneficial for understanding the content of images of a domain.
\end{abstract}
\begin{keywords}
Attention, convolutional neural networks, medical image classification
\end{keywords}

\section{Introduction}
\subsection{Background and Motivation}
Medical imaging, such as computed tomography (CT), magnetic resonance (MR), and fundus photography, has played a crucial role in the detection, diagnosis, and treatment of diseases \cite{brody2013}. Due to variations in pathology and increased size of images, researchers and doctors have been resorting to artificial intelligence (AI) techniques for automated image analysis. Among all the AI techniques, deep learning, particularly convolutional neural network (CNN), has brought the most promising performance \cite{brody2013, litjens2017, shen2017}. The successful application of CNNs is due to their automatic extraction of good feature representations which are the key to success of AI techniques \cite{bengio2013}. 

Recently, spatial attention has been introduced to CNN to improve both its performance and interpretability \cite{jaderberg2015, zhou2016, wang2017, jetley2018, schlemper2019}. The essence of spatial attention is to learn a weight map which represents relative importance of features within the same layer or channel. By promoting important features and suppressing unimportant features, the performance can be reasonably improved. In terms of interpretability, by looking at the regions to which large weights are assigned, i.e., attended regions, we may get to understand what features are mostly utilized by the CNN to produce a particular result.

All existing attention mechanisms are local attentions in the sense the weight maps are image-specific. This is because the weight map for each image is mainly determined by itself and each image is unique in the intensities of pixels. But in the medical field, even though individual images within a set are also unique, there are cases that the set of images should share the same weight map, i.e., one global weight map. The global weight map may be needed when the set of images are took on one kind of object for the same kind of disease by the same imaging technique at the same angle from which the object faces the image-taking device. We can easily tell that there exist not a few medical image datasets fit in with the procedure of the image generation mentioned above. We give a name, {\em \textbf{structured images}}, to these kinds of image sets. As the name suggests, it is expected that the structured images have organized information which conforms to a certain format. Therefore, images in other domains with organized information may be structured images as well.

When applying existing local attention mechanisms to the structured medical images, both the performance and the interpretability of a CNN may even be deteriorated. Local weight maps may cause additional overfitting since the maps are additional parameters, especially when the dataset is small-scale, which is often the case in the medical field. Local weight maps may also make the interpretation confusing because they may assign different weights to the same structured regions due to small variations in the pixel intensities. 

In this paper, we thus propose a novel global spatial attention mechanism to learn a global weight map for structured medical images. A single weight map may be ignored in terms of the number of parameters, but can still enjoy the improvement in both the performance and the interpretability.

\subsection{Novelty and Significance}
The novelty and significance of this paper is summarized as follows.

(1) {\em Proposal of a global spatial attention mechanism for the first time.} To the best of our knowledge, all existing attention mechanisms are local. Moreover, existing ones learn local weight maps in the hidden layers.
In the proposed global attention mechanism, there is only one global weight map, and the global weight map is directly learned for the input layer, i.e., the input raw images. In brief, the global weight map is instantiated by a decision boundary between important pixels and unimportant pixels. And we propose to realize the decision boundary by a binary classifier. Note that we in fact do not know which pixels are important or unimportant. Therefore, the binary classifier is integrated into the image classification CNN and to be optimized together with the CNN.

(2) {\em Applications to medical image classification.} The proposed global attention mechanism is a generic approach and can be straightforwardly integrated into any CNN architectures. We study four powerful CNNs which are GoogleNet, VGG, ResNet and DenseNet, and apply them to two medical datasets. Moreover, an additional facial expression dataset is studied because the proposed method is essentially designed for structured images. The results showed that with the proposed attention, not only the performance of GoogleNet, VGG, ResNet, and DenseNet can be improved, but also meaningful attended regions can be obtained, which is beneficial for understanding the content of images of a domain.

\subsection{Organization of This Paper}
The rest of this paper is organized as follows. Section 2 presents notations and definitions used in this paper. The proposed method is developed in section 3. Section 4 presents results of empirical evaluation. Section 5 presents related work. In section 6, conclusions are drawn from this study and future directions are introduced.

\section{Notations and Definitions}
This paper deals with structured images defined as follows:

\begin{defn} {(\textsc{Structured images}) \it Structured images have the same sizes in all dimensions, and are taken on one kind of objects under similar circumstances which include similar image-taking devices, similar context within which the objects locate, and similar angles from which the objects face the devices.}\end{defn}

\begin{figure}[t]   
		\centering
		\begin{minipage}{2cm} 
			\includegraphics[width=1.5cm, height=0.5cm]{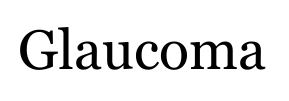}   
		\end{minipage} 
		\begin{minipage}{2cm} 
			\includegraphics[width=2cm, height=2.03cm]{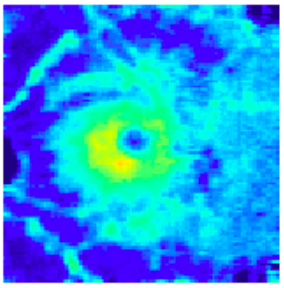}   
		\end{minipage}  
		\begin{minipage}{2cm} 
			\includegraphics[width=2cm, height=2.03cm]{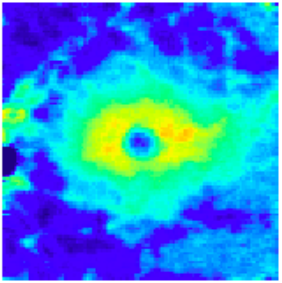}  
		\end{minipage}
		\begin{minipage}{2cm} 
			\includegraphics[width=2cm, height=2.03cm]{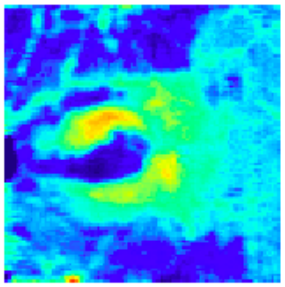}  
		\end{minipage}  
		\vspace{0.2cm}
		
		\begin{minipage}{2cm} 
			\includegraphics[width=1.1cm, height=0.5cm]{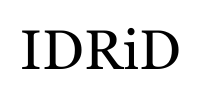}   
		\end{minipage} 
		\begin{minipage}{2cm} 
			\includegraphics[width=2cm, height=1.93cm]{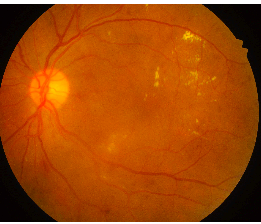}   
		\end{minipage}  
		\begin{minipage}{2cm} 
			\includegraphics[width=2cm, height=1.93cm]{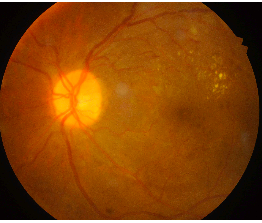}  
		\end{minipage}
		\begin{minipage}{2cm} 
			\includegraphics[width=2cm, height=1.93cm]{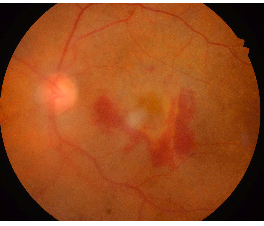}  
		\end{minipage}  
		\vspace{0.2cm}
		
		\begin{minipage}{2cm} 
			\includegraphics[width=1.1cm, height=0.5cm]{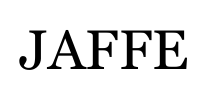}   
		\end{minipage} 
		\begin{minipage}{2cm} 
			\includegraphics[width=2cm, height=2.03cm]{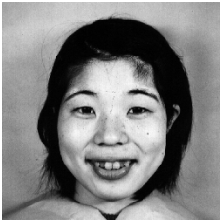}   
		\end{minipage}  
		\begin{minipage}{2cm} 
			\includegraphics[width=2cm, height=2.03cm]{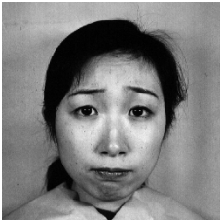}  
		\end{minipage}
		\begin{minipage}{2cm} 
			\includegraphics[width=2cm, height=2.03cm]{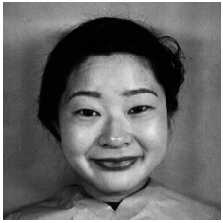}  
		\end{minipage}  
		\caption{Examples from three structured image datasets.} 
		\label{fig:example}
\end{figure}

Although the proposed method is mainly for medical image classification, it essentially works for structured images defined above. It happens that structured images are mainly produced in the medical field because of the circumstances required for the images.

The way the structured images are generated guarantees that different images have similar characteristics at the same pixels, which gives the opportunity to learn a single weight map shared by all images within a set. Three images from each of the three studied structured image datasets are presented in Fig. \ref{fig:example} for illustration. The first two datasets contain medical images taken on human eyes. We studied an additional dataset about facial expression images which focus on human faces only to demonstrate the potential applications to fields other than the medical field. Details about the description of the three datasets can be found in the following section of experiments.

The classification of images is based on their characteristics which are pixel intensities. Similarly, for the global attention weight map, we propose to learn the importance of pixels based on pixels' characteristics. In our problem setting, we only have images but have no additional information about each pixel. We thus obtain pixel characteristics by utilizing images themselves. We have learned that to measure the importance of features in the filter-style feature selection methods \cite{molina2002, chandrashekar2014, tang2014}, the characteristics of each feature can be obtained as the values of all data instances at the feature. Since the studied images are structured images, we can also obtain the characteristics of a pixel as the intensities of all images at the pixel. And it is expected that there are different distributions of intensities among different pixels. To facilitate the image classification and the pixel classification, we define an image representation and a pixel representation of a structured image dataset as follows.

\begin{figure}[t] 
   \centering
   \includegraphics[width=6.1in]{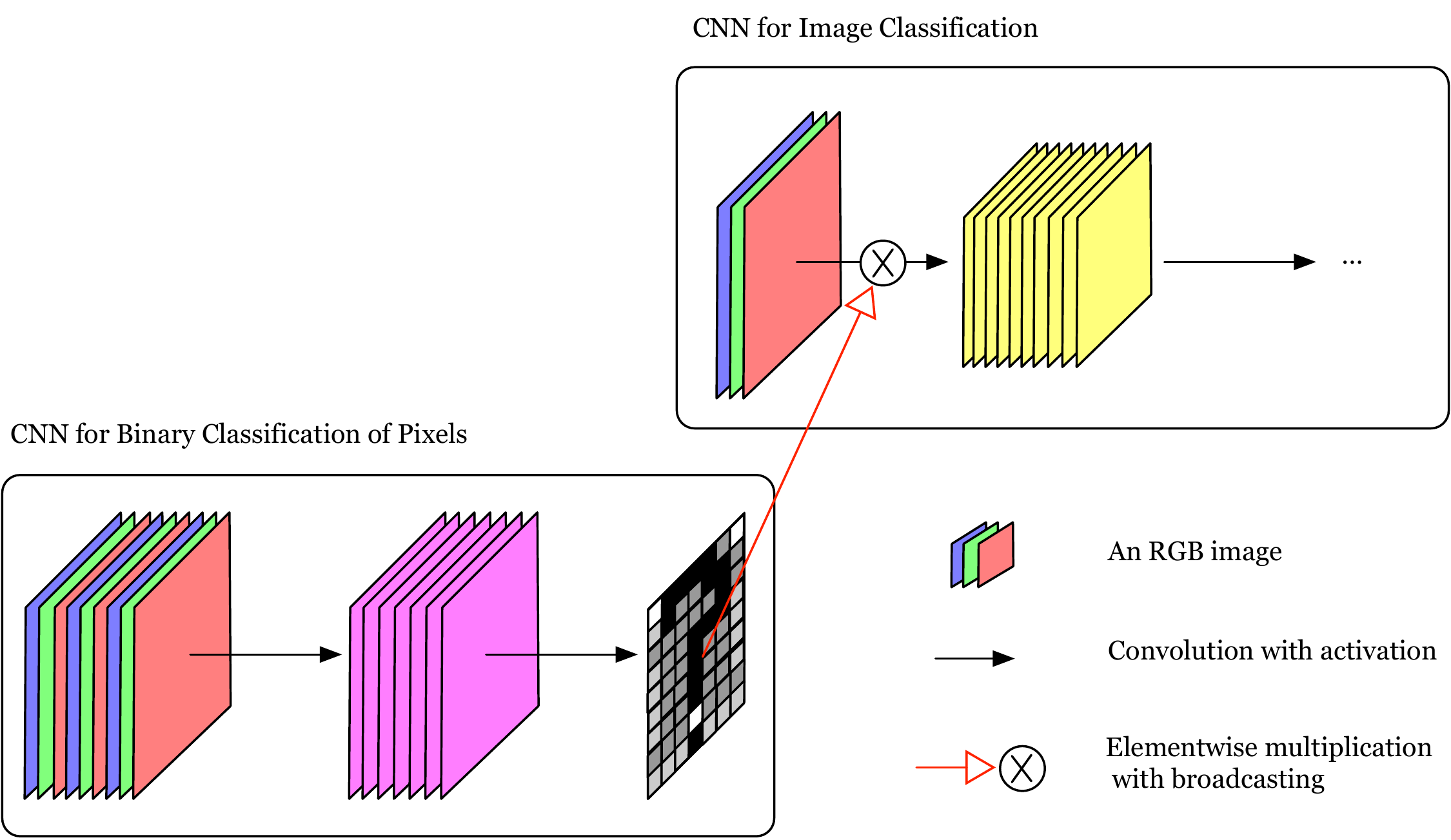} 
   \caption{Illustration of the proposed model on a toy dataset consisting of three images. The outputs of the CNN for pixel classification range from zero (white color) to one (black color).}
   \vspace{0.2cm}
   \label{fig:model}
\end{figure}

\begin{defn} {(\textsc{Image representation}) \it Images are represented by $\mathcal{X}\in \mathbb{R}^{N\times C \times W \times H}$, which is a four-dimensional tensor where $N$ is the number of images, $C$ is the number of image channels, $W$ and $H$ are the width and height of each image, respectively. } \end{defn}

\begin{defn} {(\textsc{Pixel representation}) \it Pixels are represented by $\boldsymbol P\in \mathbb{R}^{NC\times W \times H}$, which is a three-dimensional tensor obtained by reshaping $\mathcal{X}$. The number of pixels is $WH$, and the dimension of the representation of each pixel is $NC$.} \end{defn}

Note that we do not have ground-truths to perform the classification of pixels. Since the pixel importance learning is to serve the task of image classification, we thus propose to integrate the pixel classifier into the image classification CNN, and to optimize the pixel classifier together with the CNN. We define the studied problem as follows:

\begin{defn} {(\textsc{The Studied Problem}) \it Given image representation $\mathcal{X}$ and pixel representation $\boldsymbol P$, and image labels $\boldsymbol Y \in \mathbb{R}^{N}$, the objective to learn a function mapping $\mathcal{X}$ to $\boldsymbol Y$ where the function is parameterized by a conventional neural network and a pixel classifier.} \end{defn}

\section{Methodology}
\subsection{Overview}
The proposed model is illustrated in Fig. \ref{fig:model}. The model consists of two convolutional neural networks, one for the classification of images and another for the classification of pixels. In the rest of the paper, the two CNNs are referred to as image CNN and pixel CNN, respectively. There exist many powerful image CNNs such as VGG \cite{simonyan2014} and ResNet \cite{he2016}, and hence we directly employ an existing one. The novelty of the proposed model lies in the integration of the pixel CNN into the image CNN as a global spatial attention. The pixel CNN is carefully designed and described in the following subsection. The model takes two representations from a structured image dataset as input, i.e., the image representation $\mathcal{X}$ and the pixel representation $\boldsymbol P$. The pixel representation is fed to the pixel CNN to obtain the attention weight map. In particular, the pixel CNN produces the importance values of pixels which range from zero to one. The value of one and zero denote the highest importance and the lowest importance, respectively. The image representation multiplied by the importance values of pixels is fed to the image CNN as input.

\subsection{Pixel CNN}
The pixel CNN functions as a binary classifier. Since the features of a pixel are well-defined, many existing binary classifiers are applicable, such as the logistic regression and the multilayer perceptron. But instead of only the features of the pixel itself, we propose to further consider the features of surrounding pixels as its auxiliary features because nearby pixels have dependencies. Moreover, considering surrounding pixels allows for small transformations of the objects in the images and the small transformations actually may be inevitable even in the structured images. Therefore, we propose to employ a CNN which is designed to capture spatial dependencies among pixels.

\begin{figure}[t] 
   \centering
   \includegraphics[width=3.0in]{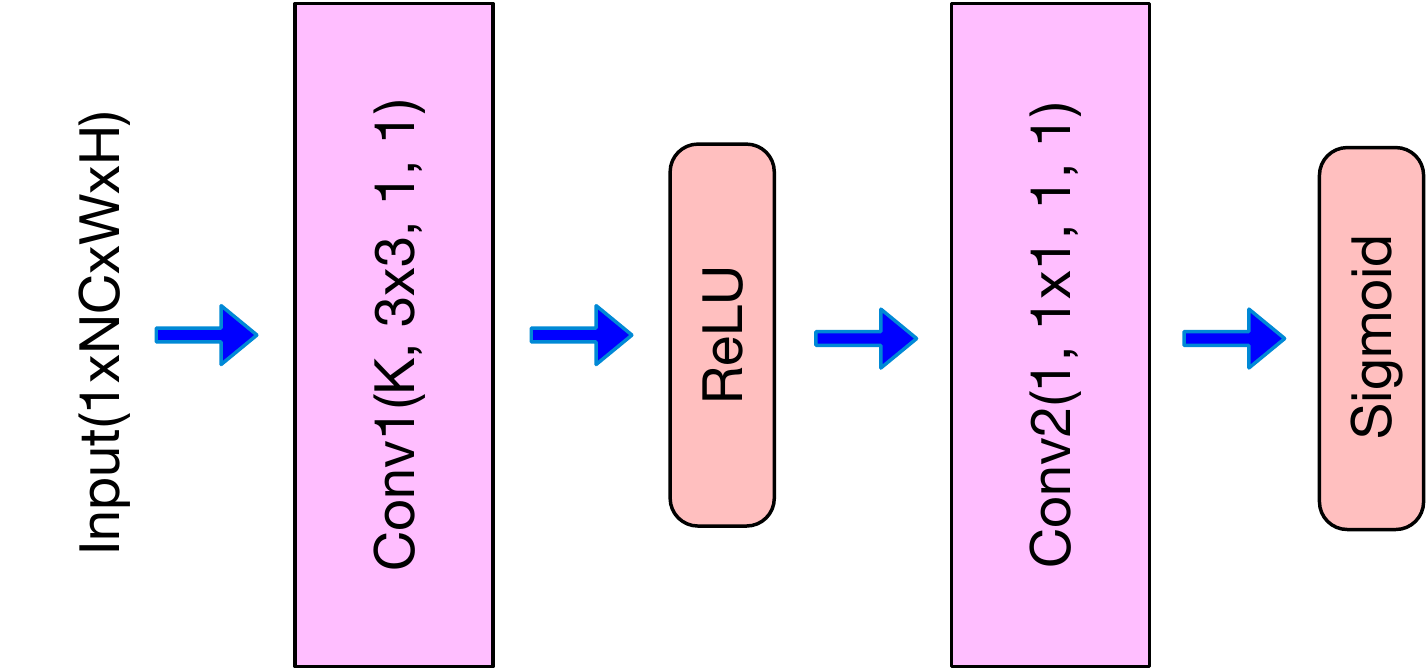} 
   \caption{Illustration of the pixel CNN where the numbers in the convolution operation are the number of channels $K$, kernel size, stride and padding, respectively.}
   \vspace{0.2cm}
   \label{fig:pixel_cnn}
\end{figure}

But different from the image CNN which can take a batch of images as input at each time, the pixel CNN should take all the pixels simultaneously as input because the importance of all the pixels should be simultaneously considered in the image CNN. The architecture of the proposed pixel CNN is illustrated in Fig. \ref{fig:pixel_cnn}. To accord with the convention of the convolution operation, the dimension of the input is denoted as $1\times NC\times W \times H$. The pixel CNN only has two convolutional layers, and works well in the experiments. Moreover, the experiments showed that the increase of hidden layers does not consistently bring improvement in performance. 

In the first convolutional layer, the design of the kernel size is to take into account surrounding pixels as auxiliary features. The stride and padding are designed to keep the number of pixels the same in the output of the convolutional layer. As a result, the first convolutional layer makes abstract features of each pixel out of its raw features and surrounding pixels' raw features. The second convolutional layer is designed to generate the importance value of each pixel by only looking at its own abstract features, which is realized by a kernel of size $1\times 1$. 

\subsection{Cost Function}
The cost function can be obtained as follows:
\begin{equation}
\mathcal{L} = \frac{1}{N}\sum_{i=1}^N \text{loss} (F(\mathcal{X}_i \ast M(\boldsymbol P;\theta_M);\theta_F), \boldsymbol Y_i) + \lambda |M(\boldsymbol P;\theta_M)|,
\end{equation}
where $\text{loss}(\cdot)$ is the cross entropy loss function, $F(\cdot;\theta_F)$ and $M(\cdot;\theta_M)$ are classification functions parameterized by the image CNN and the pixel CNN, respectively, $\ast$ realizes the element-wise multiplication with broadcasting of image pixels and corresponding pixel importance values, $\theta_F$ and $\theta_M$ are parameters of the image CNN and the pixel CNN, respectively. $\mathcal{X}_i$ is the $i$-th image, and $\boldsymbol Y_i$ is the corresponding class label. The L1 regularization on the output of the pixel CNN is to enforce pixel selection, and works well as demonstrated in the experiments. $\lambda \in \mathbb{R}$ is the coefficient.  

\subsection{Optimization}
Both the image CNN and the pixel CNN can be solved by backpropagation. And by regarding the pixel CNN as just another layer of the image CNN, all the optimization algorithms developed for CNNs can be directly utilized to solve the proposed model. We present the optimization procedure in Algorithm \ref{algorithm}.

\begin{algorithm}[t]
		\SetKwInOut{Input}{Input}\SetKwInOut{Output}{Output}
		\Input{$\mathcal{X}$, $\boldsymbol Y$, $\boldsymbol P$, $K$, $\lambda$ and $E$}
		\BlankLine
		Initialize a pre-trained CNN which is existing for image classification\;
		\For{epoch = 1, 2, 3, ..., E}{
		Optimize the image CNN and the pixel CNN\;
		}
		\For{epoch = E+1, E+2, E+3, ... }{
		Fix the pixel CNN, and optimize the image CNN only\;
		}
		\caption{The optimization procedure}
		\label{algorithm}  
\end{algorithm}

There are mainly three hyper-parameters introduced by the proposed algorithm. $K$ is the number of channels in the hidden layer, $\lambda \in \mathbb{R}$ is the coefficient of the L1 regularization on the output of the pixel CNN, and $E$ is a cut-off training epoch after which the pixel CNN is fixed. The insight behind the cut-off epoch is that after important pixels are selected, the image CNN should be fine-tuned on the important pixels. In fact, the idea of cut-off epoch is borrowed from pruning low-weight connections \cite{han2015} in deep neural networks to obtain efficient networks. In particular, the optimization procedure in the connection pruning starts with normal training, and then prunes low-weight connections among neurons, and finally fine-tunes the network on the remaining connections. Here, we firstly learn the important pixels, and fine-tune the image CNN on the learned important pixels. Note that we do not set the small importance values of pixels to zeros because it is non-trivial to determine the threshold. The cut-off epoch may depend on datasets, but works well with a small number, e.g., 60, in the experiments.

\section{Empirical Evaluation}
\subsection{Datasets}
We studied three datasets including two medical datasets and one facial expression dataset, which are described as follows:
\begin{itemize}
\item \textbf{Glaucoma} \cite{uesaka2017}: Glaucoma is an eye disease, and is the second leading cause of blindness over the world. We studied a dataset consisting of 86 eyes of 43 normative subjects and 505 eyes of 304 patients with open glaucoma (OAG). Each of the eyes belongs to one of three severity levels of the disease according the mean deviation (MD) of visual field sensitivity (Humphrey 10-2 test), i.e., early phase (MD $>$ -6 dB), moderate phase (-12 dB $<$ MD $<$ -6 db), and serious phase (MD $<$ - 12 dB). The images are retinal layers thickness produced by optical coherence tomography (OCT). The images were obtained from Tokyo University Hospital, Osaka University Hospital, Hospital of Kyoto Prefectural University of Medicine, Oike-Ikeda Eye Clinic, Shimane University Hospital and Hiroshima Memorial Hospital.

\begin{table}[t]
   \centering
   \caption{Statistics of datasets.}
   \begin{tabular}{lcl} 
       Name   & $\#$ Data Points & $\#$ Classes\\
      \midrule
      Glaucoma     & 591 & 3 \\
      IDRiD          & 516   &  3 (diabetic retinopathy) and 5 (macular edema) \\
      JAFFE       & 213 & 6 \\
      \hline
   \end{tabular}
   \label{tab:data}
\end{table}

\item \textbf{IDRiD} \cite{porwal2018}: This dataset consists of retinal fundus photographs that may have diabetic retinopathy lesions. There are two tasks corresponding to the severity level of diabetic retinopathy and diabetic macular edema, respectively. The images have $4288 \times 2848$ pixels, and have background pixels on both the left side and the right side. After removing the background pixels, we resize them into $224 \times 224$. Moreover, we horizontally flipped the images of the right eyes to reconcile the horizontal symmetry. This dataset is publicly available at \url{https://idrid.grand-challenge.org/Data/}.

\item \textbf{The Japanese Female Facial Expression (JAFFE) Database} \cite{lyons1998}: The dataset contains 213 images of 6 facial expressions posed by 10 Japanese female models. Each image has been rated on 6 emotion adjectives by 60 Japanese subjects such that each image has an averaged value for each kind of expression. We assign the emotion with the largest value as the label to each image. Original images are $256 \times 256$ gray level and were resized into $224 \times 224$ in the experiments. The images are publicly available at \url{https://zenodo.org/record/3451524#.XyI_6fj7Ts0}.

\end{itemize}

The statistics of the three datasets are summarized in Table \ref{tab:data}, and we have presented three samples from each of the datasets in Fig. \ref{fig:example}. Images from either of the first two medical datasets were taken on human eyes with the same kind of devices such that they fit in well with the definition of structured images. For the images from the facial expression dataset, even though they may not strictly accord with the definition, it is not difficult to see that different images share similar structural content at most pixels. 

The number of images of each dataset is not large, which is common in the medical field. It is obvious that for the facial images, there are many pixels irrelevant to the classification task because the task is about facial expressions. For the medical images, after learning from the medical field, we figured out that not all the pixels are relevant to the diseases as well. But the knowledge about the explicit boundary between relevant pixels and irrelevant pixels for each dataset was not provided. Therefore, CNNs may get overfitted on these image datasets with irrelevant pixels, and the proposed global attention method could be useful for alleviating the overfitting problem.

\subsection{CNN Models for Image Classification}
We studied four popular CNN models which are \textbf{VGG-16} \cite{simonyan2014}, \textbf{GoogleNet} \cite{szegedy2015},  \textbf{ResNet-152} \cite{he2016}, and \textbf{DenseNet-161} \cite{huang2017}. The four models mark four different types of powerful CNN architectures. The success of VGG is based on very small (3x3) convolution filters. Unlike VGG in which there is only one convolution with a fixed kernel size from one layer to its consecutive layer, GoogleNet utilizes multiple convolutions with different sizes. Before ResNet, many very deep neural networks suffer the problem of degradation of performance as the number of layers increases. To solve this problem, ResNet has a novel deep residual learning framework which creates an identity connection from one layer to the next layer. As a result, ResNet-152 can have as many as 152 layers and enjoys performance boosting. DenseNet further encourages the connections from early layers to later layers to strengthen feature propagation and to promote feature reuse. These CNN models are now popular building blocks for many visual tasks. We studied these different CNN models to demonstrate that the proposed global attention mechanism is a generic solution, and is beneficial for different CNN architectures in the image classification tasks.

\subsection{Baselines}
The experiments were mainly conducted to compare the performance of the CNN models with and without the proposed global attention mechanism. Since local attention mechanisms in CNNs are also available, we employed VGG with attention mechanism at the last three levels (\textbf{VGG-att3}) proposed in \cite{jetley2018} and residual attention network with 92 trunk layers (\textbf{ResAttNet-92}) proposed in \cite{wang2017} as baselines.

When the attention is regarded as a pixel selection method, it is similar to the embedded category of feature selection methods \cite{molina2002, chandrashekar2014, tang2014}. Therefore, we further compare the performance of a CNN with the attention and the same CNN with the L1 regularization. In particular, we multiply each pixel by one as the initialization weight, and enforce L1 regularization on the pixel weights. 

Moreover, because of the nature of the structured images, we studied the performance of two non-deep models, which are \textbf{logistic regression (LR)} and \textbf{support vector machine (SVM)}. We did not compare other methods making hand-crafted features because feature engineering is not needed for deep learning, e.g., the direct application of CNN on the raw retinal fundus photographs has been widely recognized \cite{gulshan2016}.

\subsection{Implementation}
For the LR and SVM, we employed the functions provided by scikit-learn \footnote{https://scikit-learn.org/stable/}. For the CNNs, we used pre-trained CNNs provided by Pytorch. For the hyper-parameters of all baseline models, we performed grid-search on their appropriate ranges. For optimizing the CNNs, we employed the Adam optimizer, and set the mini-batch size as 32, the learning rate as 0.0001, performed the search for the weight decay within the range $\{0.001, 0.0001, 0.00001\}$. For the CNNs with L1 regularization, the coefficient was searched over $\{100000, 10000, 1000, 100, 10, 1\}$ which contains large values because the regularization loss on each pixel was averaged by the number of pixels. For the proposed models, the number of channels $K$ was searched over $\{32, 62, 128, 256\}$, and the coefficient of the regularization $\lambda$ on the outputs of the pixel CNN was searched over $\{0.1, 0.01, 0.001, 0.0001\}$, and the cut-off epoch $E$ was set as 60. The total number of training epochs for all CNNs was set as 250. An example of implementation is available at \url{https://drive.google.com/drive/folders/1aNgmlXP-Xu4gsS92fmsKTxx8l4BoAx-j?usp=sharing}. The algorithm was implemented with Pytorch V1.3.1 and was executed on a GPU of GeForce RTX 2080 Ti.

\begin{table*}[t]
   \caption{Results of the image classification tasks. Bold numbers for each CNN indicate the existence of statistically significant difference between the vanilla CNN and it with global attention, and $\ast$ is for the difference between L1 and global attention.}
   \centering
   \begin{tabular}{>{\arraybackslash}m{5.1cm}>{\centering\arraybackslash}m{2.2cm}>{\centering\arraybackslash}m{2.2cm}>{\centering\arraybackslash}m{2.2cm}>{\centering\arraybackslash}m{2.2cm}}
   \toprule 
  \multirow{2}{*}{Accuracy(standard deviation)} & \multirow{2}{*}{Glaucoma} & \multicolumn{2}{c}{IDRiD} & \multirow{2}{*}{JAFFE}\\
    & & Retinopathy & Macular Edema & \\
   \midrule
   LR & 74.28 & 39.81 & 66.99 & 74.06\\
   SVM & 74.45 & 45.63 & 64.08 & 75.00\\
   VGG-att3 & 79.63(3.09) & 47.25(4.45) & 78.86(2.10) & 74.26(1.92)\\
   ResAttNet-92 & 77.17(2.33) & 41.81(5.07) & 67.15(3.98) & 70.78(6.04)\\
   \midrule
   GoogleNet & 81.06(1.18) & 52.46(4.38) & 80.49(1.90) & 78.94(2.56)\\
   GoogleNet (with L1) & 81.93(2.27) & 54.72(2.66) & 78.60(2.14) & 82.02(2.66)\\
   GoogleNet (with global attention) & \textbf{$\ast$83.61(2.43)} & \textbf{$\ast$57.21(3.20)} & \textbf{$\ast$81.36(3.05)} & \textbf{82.64(2.57)}\\
   \midrule
   VGG-16  & 82.38(1.30) & 56.96(4.63) & 80.23(2.32) & 81.78(8.44)\\
   VGG-16 (with L1) & 82.71(3.02) & 57.99(5.41) & 78.93(1.02) & 84.88(2.00)\\
   VGG-16 (with global attention) & \textbf{$\ast$84.22(1.26)} & \textbf{$\ast$66.41(1.36)} & \textbf{$\ast$83.59(0.47)} & \textbf{84.26(6.19)} \\
   \midrule
   ResNet-152  & 82.16(0.87) & 56.78(4.98) & 79.06(1.97) & 77.83(4.35)\\
   ResNet-152 (with L1) & 83.41(1.10) & 60.91(2.31) & 79.54(1.39) & 80.00(3.79)\\
   ResNet-152 (with global attention) & \textbf{$\ast$84.37(0.98)} & \textbf{$\ast$66.17(2.16)} & \textbf{$\ast$82.13(0.70)} & \textbf{80.39(3.85)}\\
   \midrule
   DenseNet-161 & 81.06(1.44) & 59.84(3.80) & 80.12(1.27) & 79.39(3.42)\\
   DenseNet-161 (with L1) & 80.08(3.18) & 60.58(1.80) & 81.07(2.56) & 82.95(3.53)\\
   DenseNet-161 (with global attention) & \textbf{$\ast$82.27(1.88)} & \textbf{$\ast$62.88(4.40)} & \textbf{$\ast$82.07(2.71)} & \textbf{$\ast$84.88(3.04)} \\
   \bottomrule
   \end{tabular}
   \label{table:results}
\end{table*}

\subsection{Evaluation Metric}
For evaluating the classification performance, we employed the metric {\em accuracy} defined as follows:
\begin{equation}
    \text{Accuracy} = \frac{\text{Number of correctly predicted images}}{\text{Total number of images}} \times 100\%
\end{equation}

\subsection{Experimental Results}
The two medical datasets were provided with 80$\%$ training instances and 20$\%$ test instances. Similarly, we separated the facial expression dataset into 80$\%$ and 20$\%$ by random. The classification accuracy is presented in Table \ref{table:results}. Since early stopping was adopted for training all the CNNs models, we designed a method based on an empirical observation to choose the epoch for measuring the performance. In particular, we performed five-fold cross validation on the training instances only, and obtained the 30 epochs which were associated with the highest accuracy as the epochs for the measurement of performance on the test instances. As a result, the resulting accuracy on the test instances by a CNN is the mean of 30 measurements with standard deviation. 

From Table \ref{table:results}, we can see that all the CNNs with the proposed global attention consistently outperformed the corresponding vanilla CNNs. We conducted the t-test between each CNN with the global attention and the vanilla CNN, and obtained the p-values less than 0.01. The reason behind the superior performance was because larger importance values were assigned to relevant pixels than to irrelevant pixels. We visualize the weight map in the following subsection. The global attention significantly outperformed the L1 regularization at most times. The exceptions happened only on the JAFFE dataset. Note that it is not difficult to distinguish the relevant pixels from the irrelevant pixels as illustrated in the face images in Fig. \ref{fig:example}. Therefore, it may not be difficult to identify the relevant pixels even for the L1 regularization. 

The poorer performance of VGG-att3 and ResAttNet compared with the vanilla VGG and ResNet (we used the pretrained version provided by Pytorch) may be due to two reasons. First, VGG-att3 and ResAttNet cannot enjoy the benefits brought by pre-training since their attention layers should be data-dependent. Second, we examined that the important pixels learned by VGG-att3 and ResAttNet for one image are always different from that of another, and figured out that the difference is due to the difference in the pixel intensities of raw images despite being structured. But in fact, all the images of a dataset should share important pixels. Therefore, the additional local attention layers may further aggravate overfitting.

\begin{figure}[t]  
\centering
\begin{tabular}{cccc}
			\textbf{a} &
			\includegraphics[keepaspectratio, width=1.55in, valign=T]{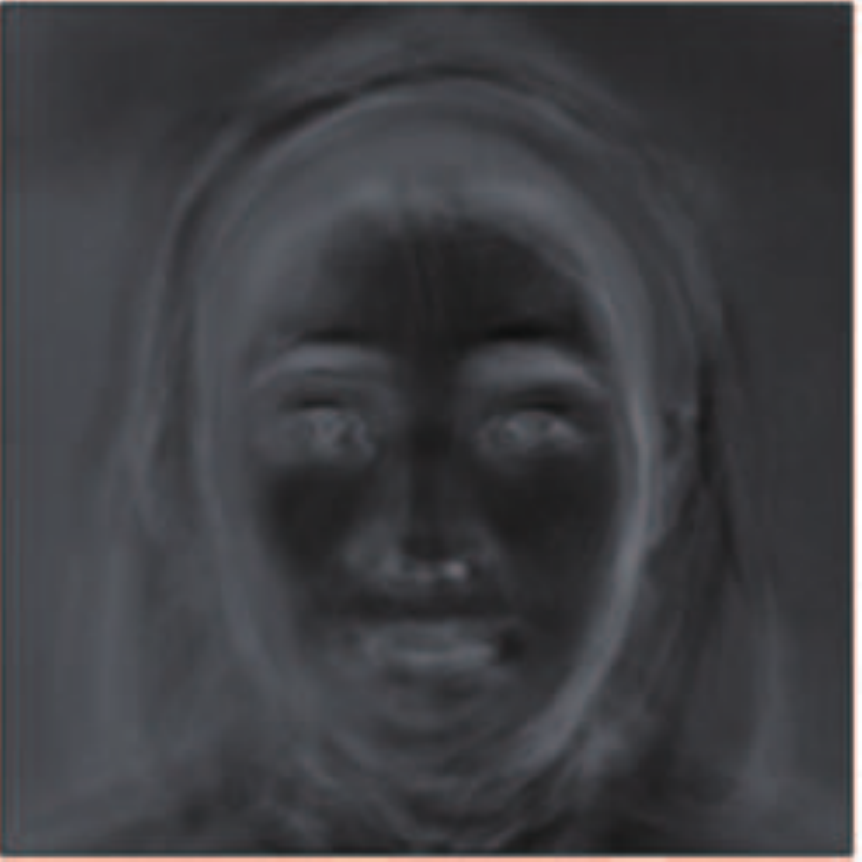}  & \textbf{b} & \includegraphics[keepaspectratio, width=1.55in, valign=T]{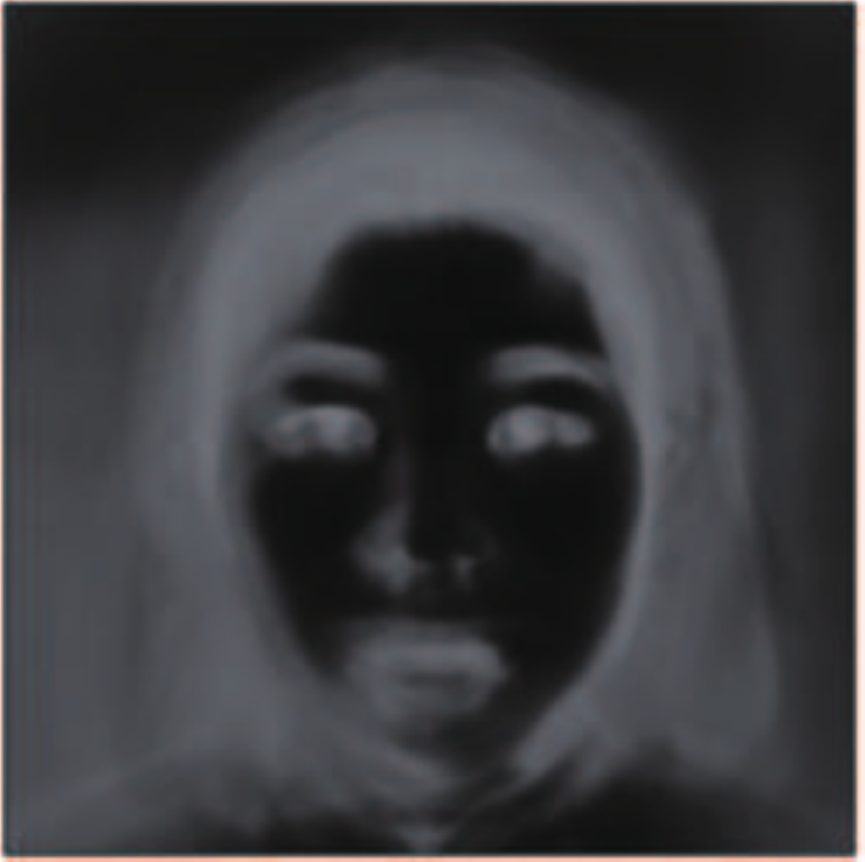}  
		\end{tabular}
		\caption{\textbf{Attention weight map for JAFFE.} \textbf{a,} attention weight map before learning. \textbf{b,} attention weight map after learning.}
   \label{fig:ROI_JAFFE}
\end{figure}

\subsection{Attention Visualization}
In this subsection, we visualized the attention weight map, i.e., the output of the pixel CNN for each dataset, and only present the representative result of one CNN for each dataset. The pixel importance values in the weight map range from zero to one. In the following visualizations, darker colors denote larger values. In fact, the smallest importance values were not always zero and the largest values were not always one. In the visualizations, the values were normalized into the range from zero to one. 

For the JAFFE dataset, we can easily tell that the ground-truth attention should be potentially paid on human faces since it is about facial expressions. The initial attention by a random initialization of the pixel CNN and the final attention after training are presented in Fig. \ref{fig:ROI_JAFFE}. It is very surprising that the initial attention appear to be consistent with the potential attention, which may be because the pixel CNN is effective at capturing spatial dependencies and common characteristics among pixels. After the learning, the attention got closer to the potential ground-truth attention. 

\begin{figure}[t]  
\centering
\begin{tabular}{cccc}
			\textbf{a} &
			\includegraphics[keepaspectratio, width=1.55in, valign=T]{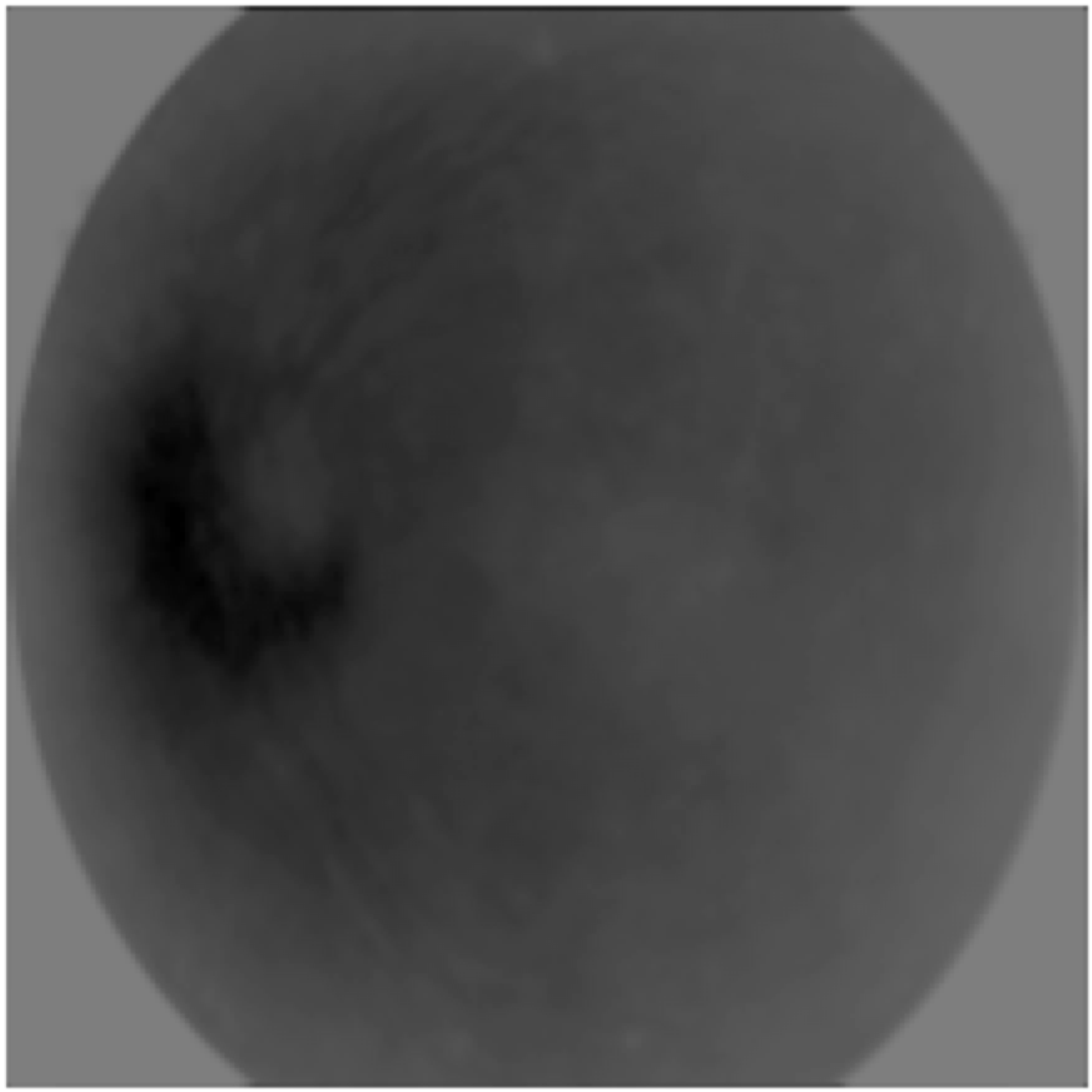}  & \textbf{b} & \includegraphics[height= 1.565in, width=1.55in, valign=T]{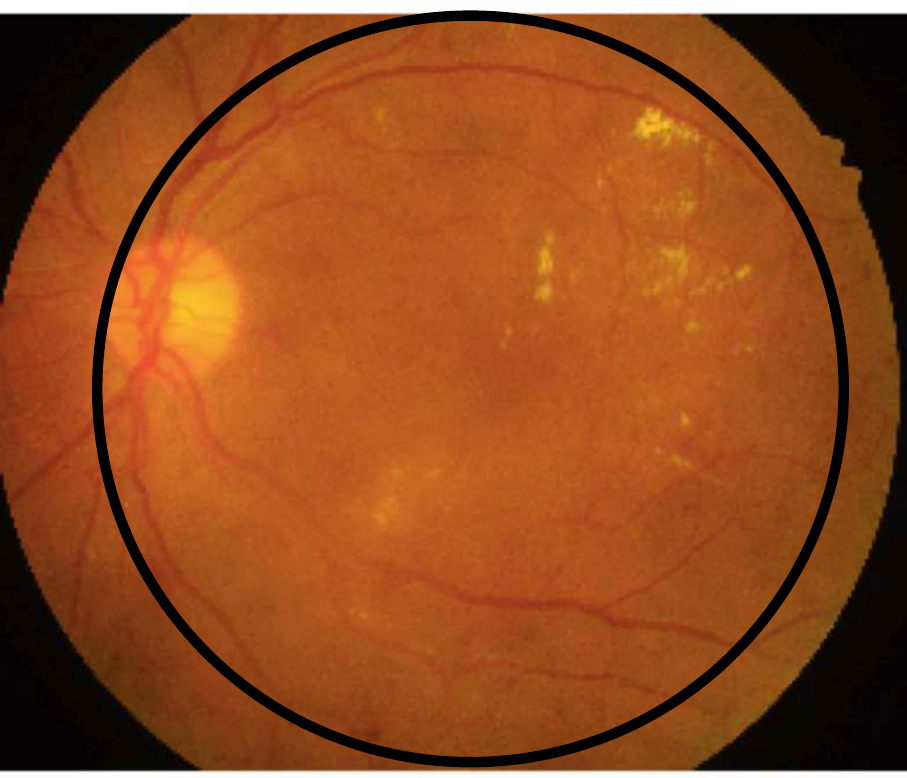}  
		\end{tabular}
		\caption{\textbf{Attention weight map for IDRID.} \textbf{a,} learned attention weight map. \textbf{b,} potential attention.}
   \label{fig:ROI_IDRiD}
\end{figure}

For the IDRiD dataset, studies \cite{deepak2011} showed that the potential attention may be a circular region centered at fovea and with one optic disc diameter as illustrated in Fig. \ref{fig:ROI_IDRiD} (b). By carefully examining the learned attention in Fig. \ref{fig:ROI_IDRiD} (a), we are able to see that there exists such a circular region out of which the color is usually lighter.

\begin{figure}[t]  
\centering
\begin{tabular}{cccc}
			\textbf{a} &
			\includegraphics[keepaspectratio, width=1.55in, valign=T]{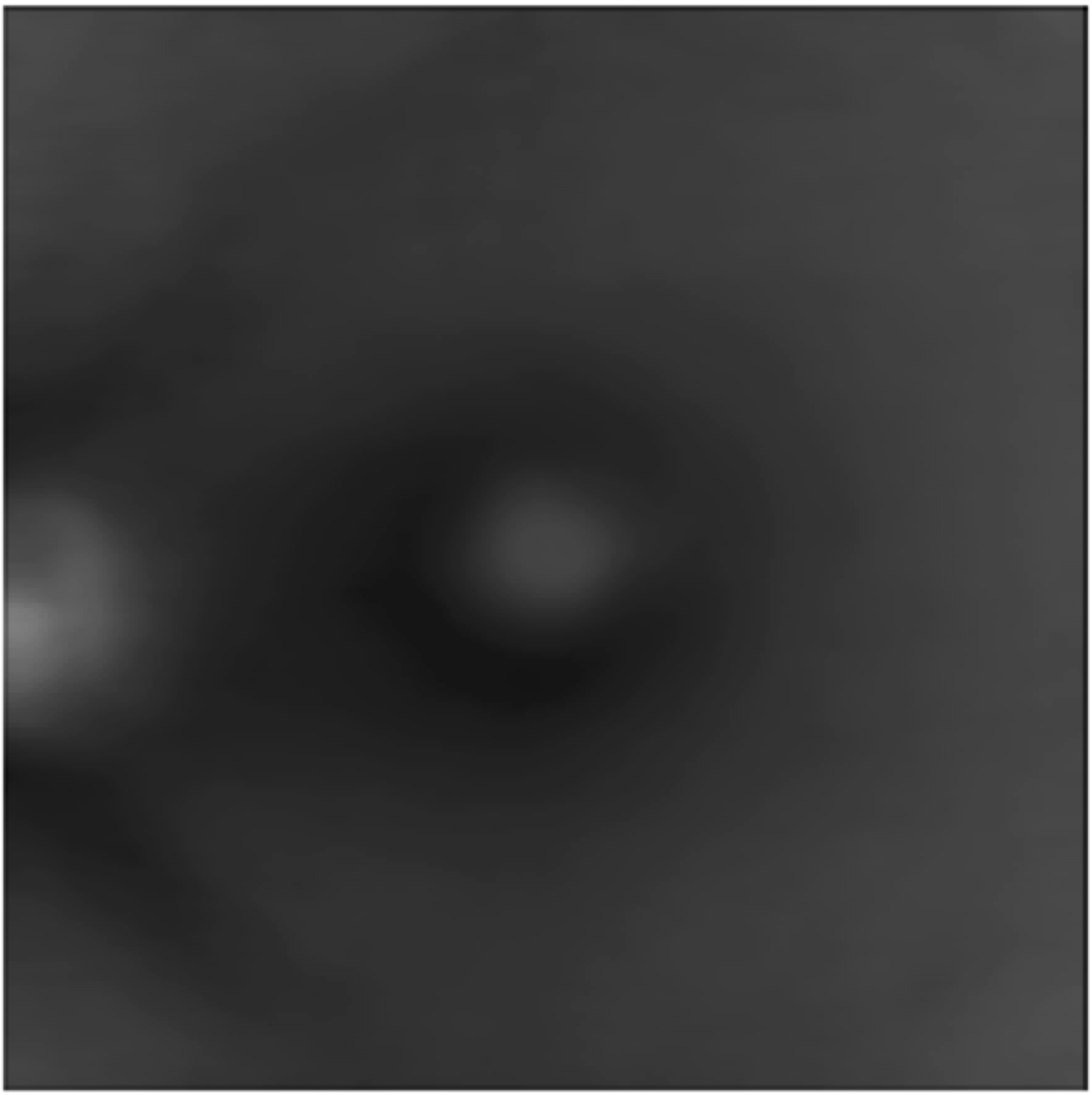}  & \textbf{b} & \includegraphics[height= 1.565in, width=1.55in, valign=T]{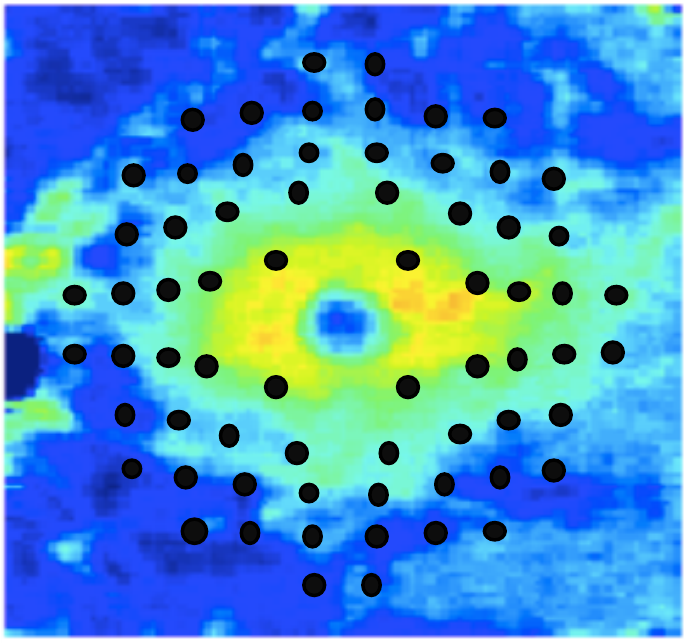}
		\end{tabular}
		\caption{\textbf{Attention weight map for Glaucoma.} \textbf{a,} learned attention weight map. \textbf{b,} potential attention.}
   \label{fig:ROI_glaucoma}
\end{figure}

For the glaucoma dataset, studies \cite{hood2013, fujino2018} showed that the potential attention locates in the region of black dots as illustrated in Fig. \ref{fig:ROI_glaucoma} (b). The dots in black are visual field (VF) test points, and the mean deviation of the test points were employed to determine the categories of the images, i.e., the severity level of glaucoma. In Fig. \ref{fig:ROI_glaucoma} (a), we are able to see that attended regions locate in the VF test points.

As a conclusion, the proposed method can learn meaningful attended regions, which is beneficial for understanding the content of images and may be relied on to guide the development of the domain knowledge.

\begin{figure}[t]  
\centering
\begin{tabular}{cccccc}
			\textbf{a} &
			\includegraphics[keepaspectratio, width=1.8in, valign=T]{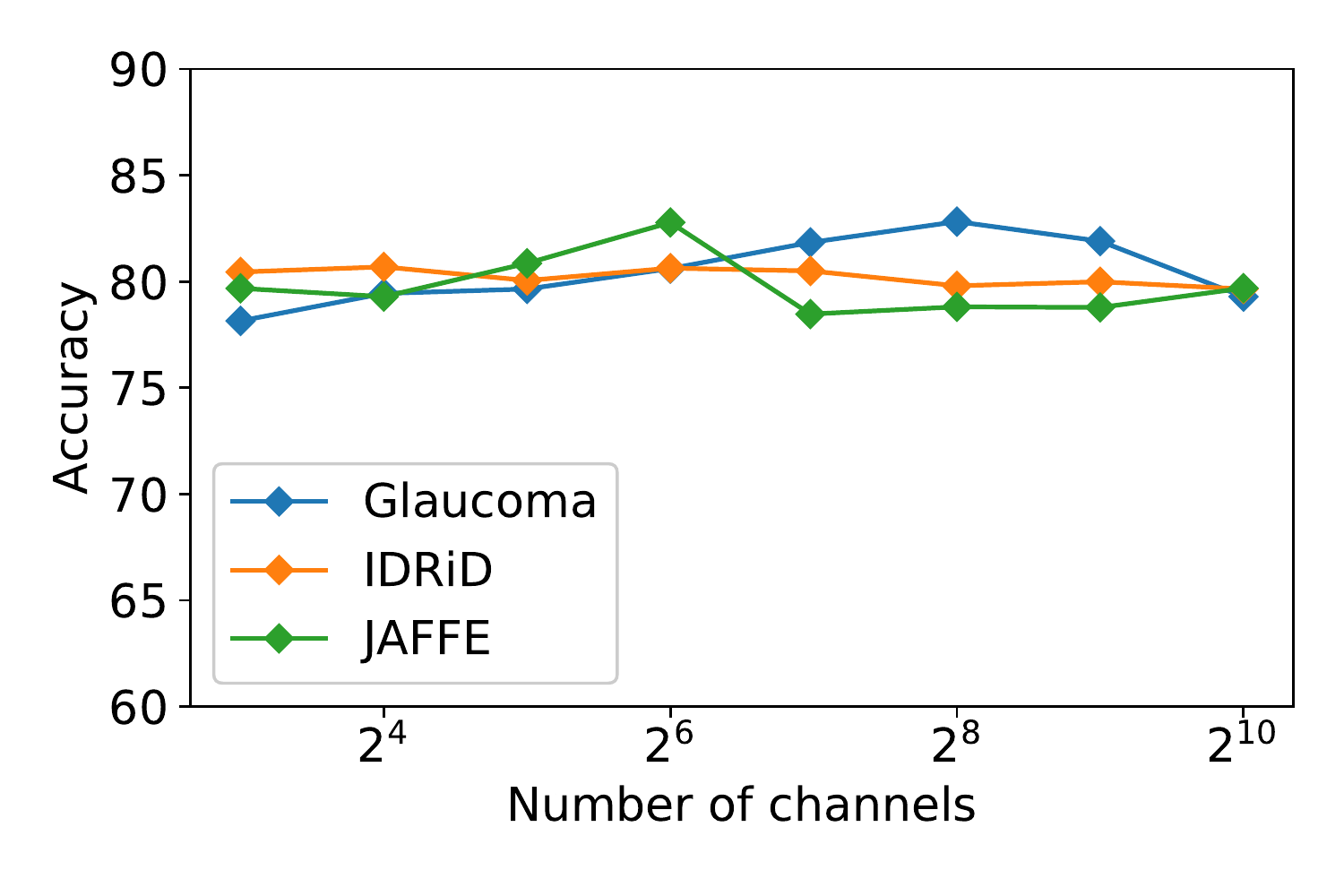}  & \textbf{b} & \includegraphics[keepaspectratio, width=1.8in, valign=T]{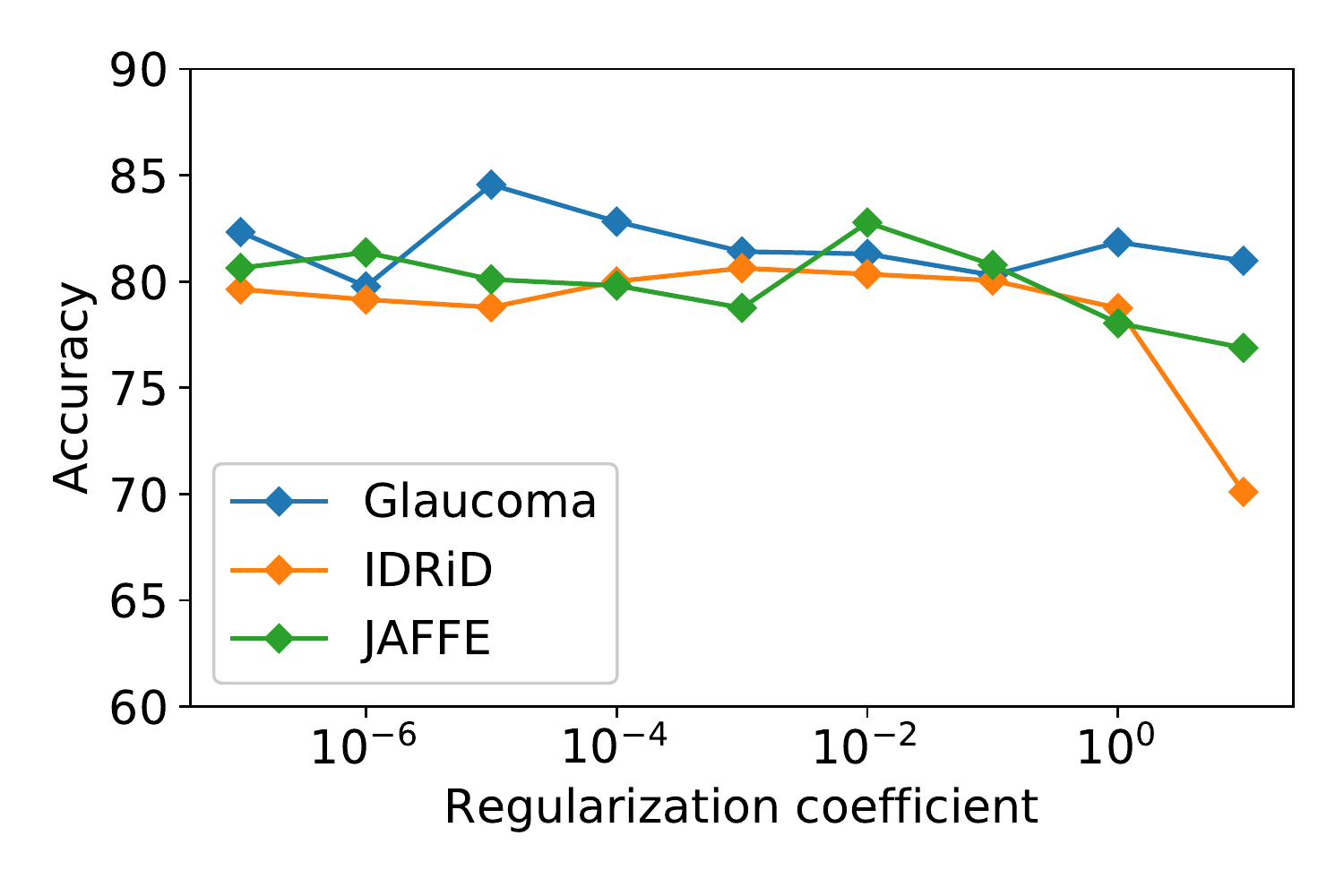}  
			\textbf{c} & \includegraphics[keepaspectratio, width=1.8in, valign=T]{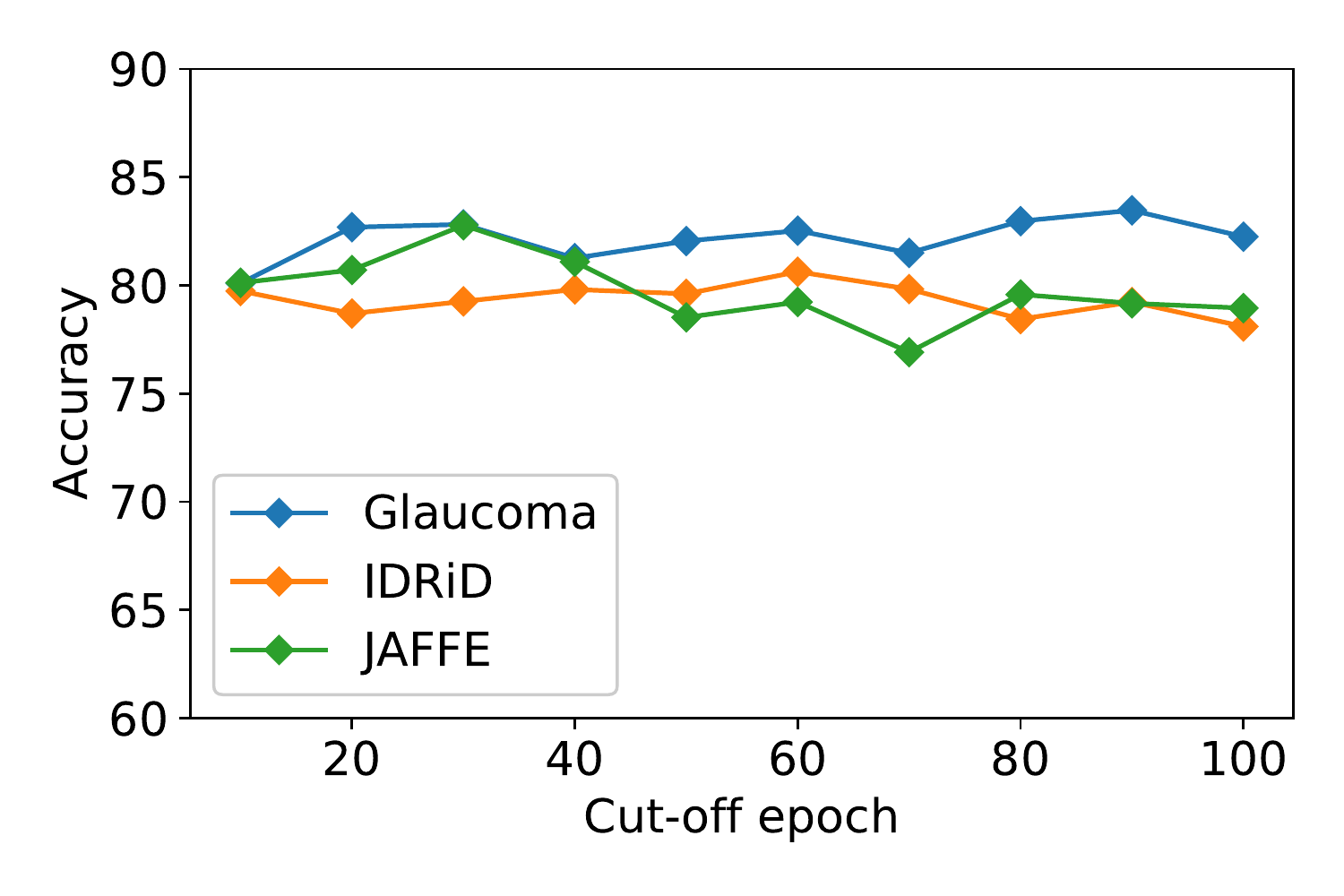} 
		\end{tabular}
		\caption{\textbf{The accuracy obtained by DenseNet when different values of hyper-parameters were adopted.} \textbf{a,} the hyper-parameter is the number of channels $K$. \textbf{b,} the hyper-parameter is the regularization coefficient $\lambda$. \textbf{c,} the hyper-parameter is the cut-off epoch $E$.}
   \label{fig:sensitivity}
\end{figure}

\subsection{Complexity Analysis}
The complexity of the proposed model is mainly determined by the number of parameters. The proposed model introduces an additional pixel CNN. The number of parameters in the pixel CNN is $K\times(3\times 3\times N\times C + 1) + (K + 1)$ where $K$ is the number of channels in the hidden layer, $N$ is the number of images, and $C$ is the number of channels of the images. The number of parameters depends on the number of images. For the three studied datasets, the number of images is not large. As a result, the number of parameters is very small compared with that of the image CNN. Even for very large datasets, such as millions of images, the number of parameters is at the same level as that of VGG, ResNet and DenseNet.

\subsection{Parameter Sensitivity}
We studied the sensitivity of the proposed model with respect to the hyper-parameters introduced by the proposed global attention mechanism, i.e., the number of channels $K$, the coefficient of the L1 regularization on the output of the pixel CNN $\lambda$, and the cut-off epoch $E$. We present the results for DenseNet in Fig. \ref{fig:sensitivity}. In the experiments, the value of a hyper-parameter was varied within an appropriate range while the other two hyper-parameters were fixed. The results show that the proposed model is not much sensitive to the three hyper-parameters within the studied range. But the coefficient for the L1 regularization should not be large (e.g., 10).

\subsection{Additional Observations on the Pixel CNN}
There actually exist many choices for designing the architecture of the pixel CNN thanks to the various architectures developed for the CNNs, and we have tried some of them. But all the other designs did not consistently bring improvement compared with the design proposed in this paper. Here, we list the studied choices as follows:
\begin{itemize}
    \item Kernel size: The kernel size in convolution determines how many surrounding pixels are considered as auxiliary features of a pixel. To keep the dimension of an image the same after the convolution, the kernel size should be odd numbers and we tried $5\times 5$ and $7\times 7$. 
    \item Number of hidden layers: In a CNN, hidden layers are important to learn abstraction of raw features. Moreover, if the kernel size is larger than $1\times 1$, the range of surrounding pixels as auxiliary features would be expanded as the number of hidden layers increases. We tried two, three, and four.
    \item Fully convolutional network: Since the last layer is implemented as a convolution layer instead of a fully connected layer, the CNN can regarded as a fully convolutional network \cite{long2015fully}. As a result, the kernel size of the last layer can be larger than one, and we tried $3\times 3$ and $5\times 5$.
    \item Dense connections: Dense connections from early layers to later layers were introduced by DenseNet.
\end{itemize}

We conducted experiments to compare the proposed design with an alternative design with respect to each individual of the four choices. For instance, while comparing the kernel with size of $3\times 3$ (the proposed) and that with $5\times 5$, we kept all the other aspects of the architecture as the same as that in the proposed architecture. The results showed that there were no statistical differences at most times between the performance obtained by the proposed design and by the alternative design in terms of each choice. As a result, we recommend the proposed architecture as the first trial if one would like to apply the proposed method. Note that the results may be data-dependent. Moreover, the combination of multiple choices may be beneficial, which is left as future studies. 

\section{Related Work}
\subsection{Medical Image Classification with CNNs}
The state-of-the-art fashion of medical image classification is to employ deep learning which can perform automatic feature extraction because good features are crucial \cite{litjens2017}. There have been many successful applications of deep learning, especially convolutional neural networks (CNNs), and several reviews are available \cite{ravi2016, shen2017, litjens2017}. The typical way of application of CNNs is to utilize existing CNN architectures which are pre-trained on natural images and then to fine-tune the pre-trained CNNs on the images at hand. The pre-training is important because powerful CNNs have excessive number of parameters and medical image datasets are usually small-scale.

However, CNNs may still get overfitted on medical images especially when there exist pixels irrelevant to the classification task, e.g., the task is only related to a particular region within the images. In order to avoid irrelevant pixels, many methods \cite{chen2015, nie2016, eppel2017, paeng2017, ni2018, zhang2018} rely on domain knowledge to obtain relevant pixels, and then somehow enforce the CNNs to pay attention to the pixels of interest. This kind of methods works at the data preprocessing stage. But the domain knowledge is not always available, and may be expensive to obtain in some scenarios. Other methods \cite{gao2015, mahapatra2016} obtain relevant pixels according to characteristics of pixel values, but still work as data preprocessing techniques.

The proposed attention mechanism can be regarded as an end-to-end data-driven approach to learning relevant pixels by estimating their importance values, and training image classification CNNs simultaneously. 

\subsection{Local Attention in CNNs}
Existing local attention mechanisms \cite{jaderberg2015, zhou2016, wang2017, jetley2018, schlemper2019} are also end-to-end data-driven approaches to learning regions of interest, but the regions of interest are image-specific. Instead, in the proposed method, the regions of interest are shared by all images within a dataset. Moreover, existing local attentions mainly work at hidden layers while the proposed global attention mechanism only works at the input layer.

\subsection{Feature Selection}
The proposed global weight map learning method can be regarded as a pixel selection method, which is similar to the concept of feature selection. Feature selection has been extensively studied to avoid the curse of dimensionality, and has been demonstrated successful on non-deep learning models. The following paragraph briefly introduces existing feature selection methods, and explains why pixel selection for CNNs has been much less studied.

Feature selection methods mainly fall into three categories, filter methods, wrapper methods, and embedded methods \cite{molina2002, chandrashekar2014, tang2014}. Filter methods are data pre-processing methods, and mainly conduct feature selection by analyzing the importance of features based on their own inherent characteristics. Since the importance of each pixel to the classification task is unknown and CNNs are able to automatically learn high-level abstraction of features, filter methods may not be preferable. Wrapper methods select features based on evaluating the performance of models on subsets of features, and multiple runs of model training are required. Since the training of CNNs is usually computationally expensive, wrapper methods are not feasible in practice. Embedded methods, as the name suggests, feature selection mechanisms are components of the model building. The commonly used embedded methods apply regularization on model parameters, e.g., L1 and L2 regularization. L1 and L2 regularization have already been employed in CNNs. But because of the huge number of parameters in CNNs, CNNs may still easily get overfitted, which explains why other regularization techniques are especially designed for deep learning models, such as dropout \cite{srivastava2014}. 

The proposed pixel selection method belongs to the embedded method, but is not another kind of regularization. Moreover, the proposed method is based on the importance of pixels to tasks where the importance is determined by inherent characteristics of a pixel. In this sense, it is similar to filter methods. But filter methods usually employ pre-defined measurements of importance, such as mutual information \cite{estevez2009} and Fisher score \cite{gu2011}. The proposed method instead measures the importance via a pure data-driven approach.

\section{Conclusion and Future Directions}
This paper for the first time proposes a global spatial attention mechanism in convolutional neural networks (CNNs) for structured medical image classification. The attention learning is realized by a binary classifier where the intensities of all images at a pixel are employed as the features of the pixel. The attention is a generic solution and can be integrated into any CNN architectures. With the global attention, the overfitting problem of CNNs can be alleviated. As a result, the experiments showed that all the studied CNNs with the proposed attention significantly outperformed the vanilla CNNs. Moreover, the attended regions are useful for understanding the structural content of the images.


The proposed method is a pure data-driven approach. In fact, for each of the studied medical datasets, the medical field has accumulated years of domain knowledge. In future, we plan to integrate the domain knowledge into the attention mechanism to further improve the performance for a particular medical dataset.

\section*{Acknowledgments}
This work was partially supported by JST KAKENHI 191400000190 and JST-AIP JPMJCR19U4. Atsushi Nitanda was partially supported by JSPS Kakenhi (19K20337) and JST-PRESTO.

\section*{Declarations of interest}
The authors declare no competing interests.

\end{document}


\setlength{\abovedisplayskip}{2pt}
\setlength{\belowdisplayskip}{2pt}

\maketitle

Section 1 presents data preprocessing. Section 2 presents the implementation.

\section{Preprocessing of Datasets}
\subsection{IDRiD}
This dataset is publicly available at \url{https://idrid.grand-challenge.org/Data/}. For each image, background pixels on both the left side (pixel index on the horizontal axis less than 260) and the right side (pixel index larger than 3685) were removed. Afterwards, we resized them into $224\times224$ using $interpolation=cv2.INTER\_AREA$ in Python. Moreover, we horizontally flipped the images of the right eyes to reconcile the horizontal symmetry. For information, the set of flipped images in the training data by index is $\{$0, 6, 8, 9, 12, 13, 16, 17, 19, 21, 22, 24, 26, 28, 30, 32, 34, 36, 38, 40, 45, 46, 48, 51, 54, 
56, 58, 60, 61, 63, 64, 65, 66, 68, 70, 73, 75, 77, 79, 81, 84, 86, 89, 90, 93, 94, 98, 99, 100,
101, 102, 104, 106, 109, 111, 115, 117, 118, 119, 123, 124, 127, 128, 131, 134, 135, 137, 139,
141, 144, 145, 147, 149, 151, 152, 156, 157, 158, 159, 160, 163, 166, 168, 170, 171, 172, 174,
176, 177, 181, 183, 184, 186, 187, 188, 190, 192, 193, 194, 195, 197, 200, 201, 202, 204, 206, 
209, 211, 212, 214, 216, 217, 220, 222, 226, 230, 234, 235, 238, 241, 244, 246, 248, 251, 252,
253, 256, 258, 259, 261, 264, 268, 270, 272, 273, 274, 275, 276, 279, 282, 284, 286, 288, 289,
291, 292, 299, 300, 302, 304, 305, 306, 307, 309, 312, 314, 315, 318, 320, 321, 322, 325, 329,
331, 333, 334, 337, 340, 341, 343, 345, 349, 351, 356, 357, 359, 361, 362, 367, 369, 370, 372, 
373, 375, 377, 378, 381, 383, 384, 387, 389, 392, 394, 396, 398, 402, 403, 405, 406, 407, 408,
409, 411, 412$\}$, and the set in the test data is $\{$1, 3, 4, 5, 8, 9, 14, 15, 16, 19, 21, 24, 26, 27, 31, 34, 35, 36, 38, 39, 41, 42, 45, 48, 50, 53, 55, 57, 60, 62, 63, 64, 66, 67, 69, 71, 76, 81, 82, 84, 86, 87, 90, 94, 96, 100, 102$\}$. Finally, the pixel values were normalized into range from 0 to 1, and then were standardized using means and standard deviations in the set $\{$0.485 and 0.229 for the first channel, 0.456 and 0.224 for the second channel, 0.406 and 0.225 for the third channel$\}$.

\subsection{JAFFE}
The images are publicly available at \url{https://zenodo.org/record/3451524#.XyI_6fj7Ts0}. There are six values corresponding to six emotions for each image, and we assigned the emotion with the largest value as the label to each image. Each image was resized into $224\times224$ using $interpolation=cv2.INTER\_AREA$ in Python. The normalization and standardization of the pixel values were performed in the same way as that for the IDRiD data.

\subsection{Glaucoma}
The preprocessing for the Glaucoma data is basically similar to that for the other two datasets. More details are omitted due to review policies and privacy issues.

\section{Implementation}
\subsection{The Proposed Model}
An example of the implementation of the proposed model is provided through Google Drive at \url{https://drive.google.com/drive/folders/1aNgmlXP-Xu4gsS92fmsKTxx8l4BoAx-j?usp=sharing}. 

\subsection{Image CNNs}
The image CNNs, e.g., GoogleNet, VGG-16, ResNet-152, and DenseNet-161, are pretrained models provided by Pytorch, which were instantiated from $torchvision.models$.

\subsection{CNNs with Attention}
For VGG-att3, we used the implementation publicly available at \url{https://github.com/SaoYan/LearnToPayAttention}.
For ResAttNet-92, we used the implementation publicly available at \url{https://github.com/osmr/imgclsmob}.

\subsection{SVM and LR}
We used the models implemented in $sklean$ \url{https://scikit-learn.org/stable/}.